\documentclass[prb,aps,twocolumn,superscriptaddress,showpacs]{revtex4}

\usepackage{graphicx}

\begin{document}

\title{Damped Soft Phonons and Diffuse Scattering in 40\%Pb(Mg$_{1/3}$Nb$_{2/3}$)O$_{3}$-60\%PbTiO$_{3}$}

\author{C. Stock}
\affiliation{Department of Physics, University of Toronto, Toronto, Ontario, M5S 1A7}
\affiliation{Department of Physics and Astronomy, Johns Hopkins University, Baltimore, MD, 21218}
 
\author{D. Ellis}
\affiliation{Department of Physics, University of Toronto, Toronto, Ontario, M5S 1A7}

\author{I. P. Swainson}
\affiliation{National Research Council, Chalk River, Ontario, Canada, KOJ 1JO}

\author{Guangyong Xu}
\affiliation{Physics Department, Brookhaven National Laboratory, Upton, New York, 11973}

\author{H. Hiraka}
\affiliation{Physics Department, Brookhaven National Laboratory, Upton, New York, 11973}

\author{Z. Zhong}
\affiliation{National Synchrotron Light Source, Brookhaven National Laboratory, Upton, New York, 11973}

\author{H. Luo}
\affiliation{Shanghai Institute of Ceramics, Chinese Academy of Sciences, Shanghai, China, 201800}

\author{X. Zhao}
\affiliation{Shanghai Institute of Ceramics, Chinese Academy of Sciences, Shanghai, China, 201800}

\author{D. Viehland}
\affiliation{Department of Materials Science and Engineering, Virginia Tech., Blacksburg, Virginia, 24061}

\author{R.J. Birgeneau}
\affiliation{Department of Physics, University of Toronto, Toronto, Ontario, M5S 1A7}
\affiliation{Department of Physics, University of California at Berkeley, Berkeley, CA, 94720}

\author{G. Shirane}
\affiliation{Physics Department, Brookhaven National Laboratory, Upton, New York, 11973}

\date{\today}

\begin{abstract}

Using neutron elastic and inelastic scattering and high-energy x-ray diffraction, we present a comparison of 40\% Pb(Mg$_{1/3}$Nb$_{2/3}$)O$_{3}$-60\% PbTiO$_{3}$ (PMN-60PT) with pure Pb(Mg$_{1/3}$Nb$_{2/3}$)O$_{3}$ (PMN) and PbTiO$_{3}$ (PT).  We measure the structural properties of PMN-60PT to be identical to pure PT, however, the lattice dynamics are exactly that previously found in relaxors PMN and PZN.  PMN-60PT displays a well-defined macroscopic structural transition from a cubic to tetragonal unit cell at 550 K.  The diffuse scattering is shown to be weak indicating that the structural distortion is long-range in PMN-60PT and short-range polar correlations (polar nanoregions) are not present.  Even though polar nanoregions are absent, the soft optic mode is short-lived for wavevectors near the zone-centre.  Therefore, PMN-60PT displays the same waterfall effect as prototypical relaxors PMN and PZN.  We conclude that it is random fields resulting from the intrinsic chemical disorder which is the reason for the broad transverse optic mode observed in PMN and PMN-60PT near the zone centre and not due to the formation of short-ranged polar correlations.  Through our comparison of PMN, PMN-60PT, and pure PT, we interpret the dynamic and static properties of the PMN-xPT system in terms of a random field model in which the cubic anisotropy term dominates with increasing doping of PbTiO$_{3}$.
   
\end{abstract}


\maketitle

\section{Introduction}

Relaxor ferroelectrics have generated considerable interest recently due to their promising application as piezoelectric devices.~\cite{Ye98:81,Park97:82}  Pb(Mg$_{1/3}$Nb$_{2/3}$)O$_{3}$ (PMN) and Pb(Zn$_{1/3}$Nb$_{2/3}$)O$_{3}$ (PZN) are prototypical relaxor systems which display a diffuse transition with a broad and frequency dependent peak in the dielectric response.  Despite an intense amount of research into the structural and dynamic properties of these materials, there is little consensus and understanding of the relaxor's interesting structural and dielectric properties.

In spite of the presence of a peak in the dielectric response, the bulk unit cell in both PMN and PZN remains cubic at all temperatures in the absence of an applied electric field.~\cite{Xu03:67,Stock04:69}  Even though no macroscopic structural transition occurs, strong diffuse scattering around the Bragg peaks is observed in both PMN and PZN at low temperatures.   The diffuse scattering is onset near the Burns temperature T$_{d}$, where index of refraction measurements suggest that local regions of ferroelectric order are formed in a paraelectric background.~\cite{Burns83:48} This interpretation has recently be further justified by a pair-distribution function analysis and $^{207}$Pb NMR.~\cite{Jeong05:94, Blinc03:91} These local regions of ferroelectric order have been referred to as polar nanoregions.  

The phonons in PMN and PZN are characterized by a soft, zone center, transverse optic mode (TO) which becomes highly damped between the Burns temperature T$_{d}$, where polar nanoregions are formed, and the lower critical temperature T$_{c}$.~\cite{Nab99:11}  T$_{c}$ in both PMN and PZN is defined as the temperature at which the dielectric susceptibility displays a sharp frequency independent peak under the application of an electric field.  The damping of the TO mode is restricted to a small region around the zone centre.  The interpretation of the critical wavevector where the TO mode becomes broad has been the subject of much debate recently with several models being proposed including dampening from polar nanoregions, mode-coupling, and a soft quasi-optic mode.~\cite{Hlinka03:91,Vak02:xx,Gehring01:63,Cowley05:5584}  In contrast to the optic phonons, the transverse acoustic (TA) mode does not become strongly damped, however, the linewidth of the TA mode does show a subtle broadening at T$_{d}$, and a recovery below T$_{c}$.~\cite{Wakimoto02:65}

\begin{figure}[t]
\includegraphics[width=8cm] {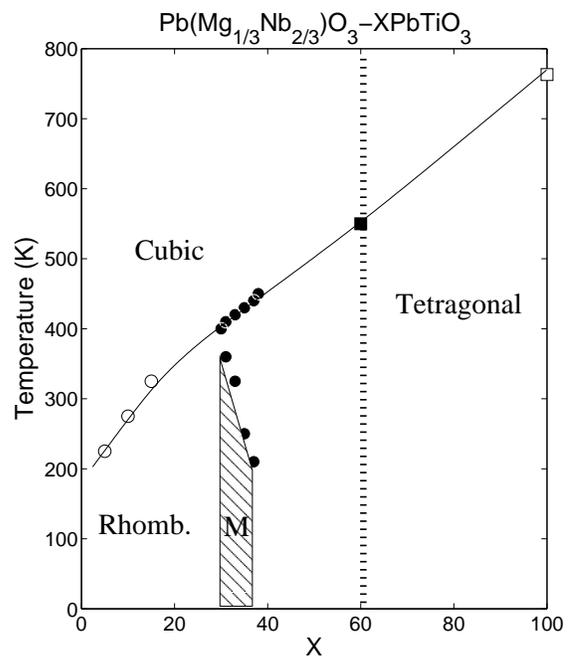}
\caption{\label{figure1}  The crystal system as a function of PbTiO$_{3}$ doping is plotted.  For low-PT concentrations, the unit cell is rhombohedral while for large PT concentrations the unit cell is tetragonal.  For intermediate concentrations, a monoclinic unit cell is observed.  The vertical dashed line marks the position of PMN-60PT in the phase diagram.  Points are obtained from Ye \textit{et al.} (open circles)~\cite{Ye03:67}, Noheda \textit{et al.}~\cite{Noheda02:66} (solid circles), and Shirane \textit{et al.} (open square).~\cite{Shirane70:2} Note that $M$ refers to a monoclinic unit cell.}
\end{figure}

Doping with PbTiO$_{3}$ (PT) in PMN or PZN has been shown to suppress the relaxor behavior, and drive the system toward a conventional ferroelectric.  The structural phase diagram as a function of PbTiO$_{3}$ doping is quite intricate and has been investigated using high-resolution synchrotron x-ray diffraction.  The phase diagram is reproduced in Fig. \ref{figure1}.  With increasing PT doping, the low temperature unit cell changes shape from cubic to rhombohedral, monoclinic, and tetragonal respectively.~\cite{Noheda02:66,Ye03:67} The dielectric response for low PT dopings is highly frequency dependent, indicating strong relaxor behavior.  For large PT dopings in the tetragonal region, the dielectric response is largely frequency independent with the response characterized by a sharp peak around the critical temperature.~\cite{Lente04:xx, Feng04:16, Ohwa01:70} Therefore, the relaxor nature, as characterized by the frequency dependence of the dielectric susceptibility, is suppressed for large PT dopings when the unit cell is tetragonal, and when the structural properties begin to strongly resemble those of pure PbTiO$_{3}$.~\cite{Shirane70:2}

Despite many studies, no unified model of the relaxor transition has been presented.  Previously, we were able to explain the dynamic and static properties off PMN and PZN in terms of a simple random field model involving two competing energy scales.~\cite{Stock04:69} This model was able to account for the two temperature scales and the lack of a clear bulk structural transition at T$_{c}$.  Other models to describe the relaxors in terms of random field and random bond models have also been proposed.~\cite{Westphal92:68,Pirc99:60,Fisch03:67}  It is clearly important to try and extend these models to describe the entire phase diagram.  In this paper, we will show that the PMN-60PT can also be interpreted in terms of random fields.  This shows that indeed the entire PMN-xPT maybe understand in terms of a simple unified picture.

We investigate the static and dynamic properties of PMN-60PT and compare them to those of PbTiO$_{3}$ and the relaxor Pb(Mg$_{1/3}$Nb$_{2/3}$)O$_{3}$ in an effort to present a unified picture for the PMN-xPT phase diagram.  The purpose of this paper is three-fold.  First, we compare the elastic scattering in PMN-60PT with PMN and PT.  We will show that PMN-60PT undergoes a well defined structural transition in a similar manner to PT.  Second, we compare the diffuse scattering in PMN-60PT to other relaxors and show that short-ranged polar correlations are absent at low temperatures.  This is indicated by the absence of temperature dependent diffuse scattering as observed in the relaxors PMN and PZN.  Third, we compare the lattice dynamics in PMN-60PT and show the soft optic-mode is very similar to PMN in that it is heavily damped near the zone centre.  The anomolous lattice dynamics near the zone-centre therefore cannot be attributed to the presence of short-range polar correlations as previously suggested.  We attribute the dampening of the zone-centre optic phonon to presence of random fields and discuss the PMN-xPT phase diagram in terms of a random field model where the cubic anisotropy term dominates with increased PT doping.

\section{Experimental Details}

Neutron elastic and inelastic measurements were conducted at the C5 and N5 triple-axis spectrometers located in the National Research Universal (NRU) reactor at Chalk River Nuclear Laboratories.  For all measurements the sample of PMN-60PT was mounted such that reflections of the form (HK0) lay within the scattering plane.  For temperatures between 100-600 K, the sample was mounted in an orange cryofurnace.  For temperatures between 600-800 K, a high-temperature furnace was used which utilized a molybdenum radiative shield to ensure that the sample was heated uniformly.  To prevent degradation of the sample, the temperature was limited to 800 K.  The $\sim$ 4 cc crystal was grown by the modified Bridgman technique previous described.~\cite{Luo00:39}  To compare the diffuse intensity located around the Bragg peaks to that of PMN, diffuse scattering measurements were carried out on a 4 cc sample of pure PMN also grown by the modified Bridgeman technique.  The sample was aligned in the (HK0) plane in an orange cryofurnace.  To compare absolute intensities with those of PMN-60PT, all diffuse scattering data were normalized to the intensity of a transverse acoustic phonon measured at Q=(2, 0.14, 0).

For measurements on C5, a variable vertically focussing pyrolytic graphite (002) monchromator and a flat graphite (002) analyzer were used with the horizontal collimation fixed at 12$'$-33$'$-S-29$'$-72$'$.  A graphite filter was used on the scattered side to filter out higher order neutrons and a liquid nitrogen cooled sapphire filter was used before the monochromator to reduce the fast neutron background.  Measurements on N5 were conducted with a flat graphite monochromator and analyzer and horizontal collimations were set to 30$'$-26$'$-S-24$'$-open.  All measurements were conducted by fixing the final energy at 14.6 meV and varying the incident energy defining the energy transfer as $\hbar \omega$=E$_{i}$-E$_{f}$.  All inelastic data was corrected for higher-order contamination of the incident beam monitor as previously described.~\cite{Shirane:book}

X-ray diffraction measurements were performed on the X17B1 beamline of the National Synchrotron Light Source (NSLS) located at Brookhaven National Laboratories.  A monochromatic x-ray beam of 75 keV with an energy resolution of 10$^{-4}$ ($\Delta E/E$) was used.  A CCD detector was put after the sample to acquire diffraction images.  The beam was incident along the [001] crystallographic direction such that the CCD recorded diffraction patterns close to the (HK0) plane.  This allowed the diffuse scattering in PMN-60PT to be mapped out over several wavevectors simultaneously.  For both PMN and PMN-60PT, the sample was aligned in a displex such that low-temperatures could be reached and thermal diffuse scattering arising from phonons could be reduced.  The method used to map out the diffuse scattering is the same used previously to measure the three-dimensional lineshape of the diffuse scattering in PZN-xPT.~\cite{Xu04:70}  

\section{Elastic Scattering}

In this section we first investigate the structural transition in PMN-60PT.  We show the absence of low-temperature diffuse scattering and hence find that polar nanoregions are not present.  We will show that there exists a weak high temperature diffuse component previously observed in pure PMN and attributed to short-range chemical order. 

\subsection{Structural Properties: First order Structural Transition}

\begin{figure}[t]
\includegraphics[width=8cm] {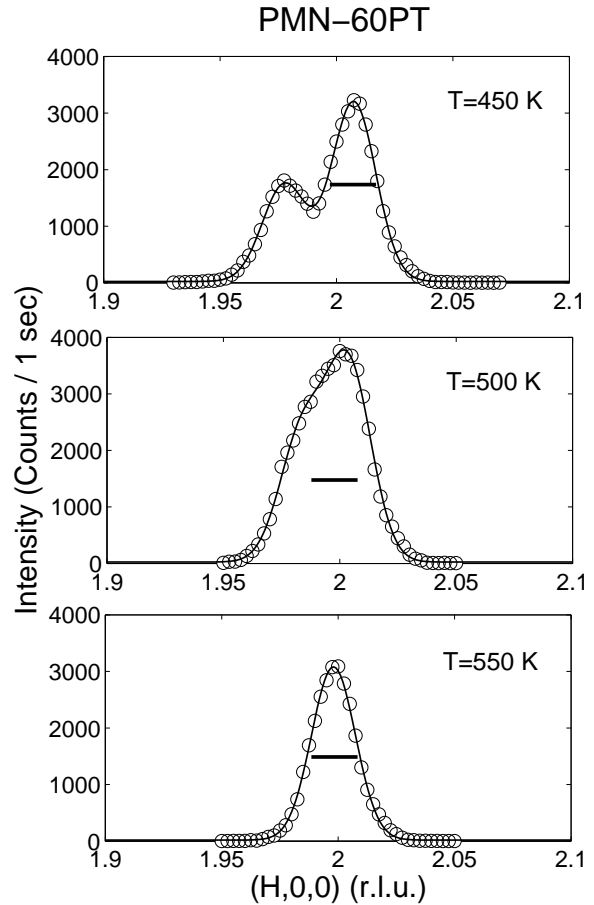}
\caption{\label{figure2} Radial scans through the (200) Bragg peak are plotted for various temperatures.  At low temperatures, the presence of two peaks indicates that the unit cell is tetragonal in shape where a single sharp peak at high temperatures indicates that the unit cell is cubic.  The resolution is given by the horizontal bar.}
\end{figure}

\begin{figure}[t]
\includegraphics[width=8cm] {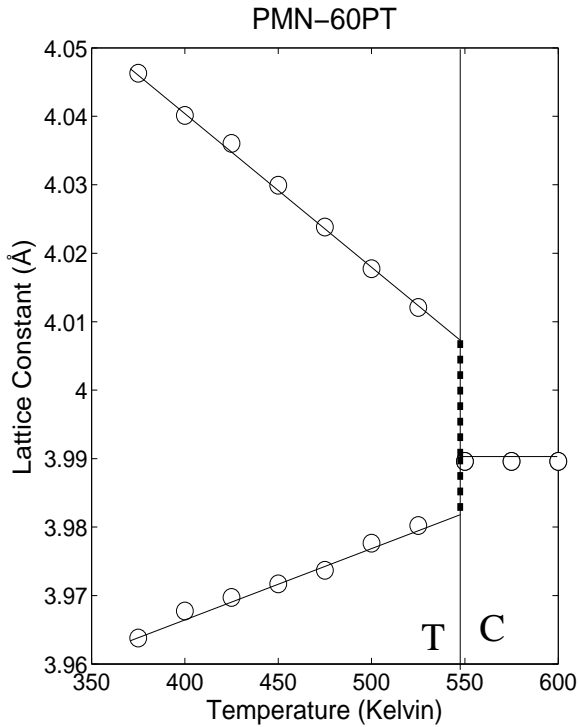}
\caption{\label{figure3} The lattice parameters as a function of temperature is plotted around the critical temperature T$_{c}$.  The two lines below T$_{c}$ correspond to the $a$ and $c$ lengths.}
\end{figure}

The structural transition in PMN-60PT was studied with neutron diffraction by conducting scans through the (200) Bragg peak.  Examples are illustrated in Fig. \ref{figure2} at 450, 500, and 550 K.  At high temperatures, a sharp, single peak is observed indicative of a cubic unit cell characterized by a single lattice parameter.  This sharp resolution limited peak splits into two peaks as a result of the unit cell changing shape from cubic to tetragonal.  The two peaks illustrate the formation of domains with different lattice parameters along the $a$ and $c$ axes.   The lattice constant as a function of temperature is plotted in Fig. \ref{figure3}, a transition from cubic to tetragonal lattice parameters is observed.  The change in lattice constant near the critical temperature, T$_{c}$, is very different from those of pure PMN where no strong anomaly in the lattice constant is observed near T$_{c}$ in single crystal samples.~\cite{Dkhil04:65}  The presence of a sharp well defined structural transition is consistent with the dielectric results, which show a frequency independent peak in the dielectric response.  This contrasts with PMN, where the peak in the dielectric constant is highly frequency dependent.  Therefore, the structural properties of PMN-60PT are very different to that of pure PMN observed from both neutron elastic scattering and dielectric measurements.

The presence of a distinct bulk transition from a cubic unit cell to a tetragonal unit cell is very similar to the cubic to tetragonal distortion observed to occur in PbTiO$_{3}$ at 750 K.  The transition (characterized by the lattice constants) is abrupt as expected for a first order transition as seen in pure PbTiO$_{3}$.~\cite{Shirane51:6}  This is very different to that observed in PMN, PZN, and low doped PMN-xPT and PZN-xPT relaxors.  In PMN and PZN, the unit cell remains cubic in shape at all temperatures, though PZN has been found to have a near surface region where a structural transition is observed.~\cite{Xu03:67,Xu04:70_1}  With increasing PT doping, the bulk unit cell in PMN does eventually undergo a structural transition, though PMN-10PT has been found to have an anomalous near surface layer similar in nature to that observed in PZN.~\cite{Gehring04:16}  

\subsection{Diffuse Scattering and High-temperature Chemical Short Range order}

\begin{figure}[t]
\includegraphics[width=8cm] {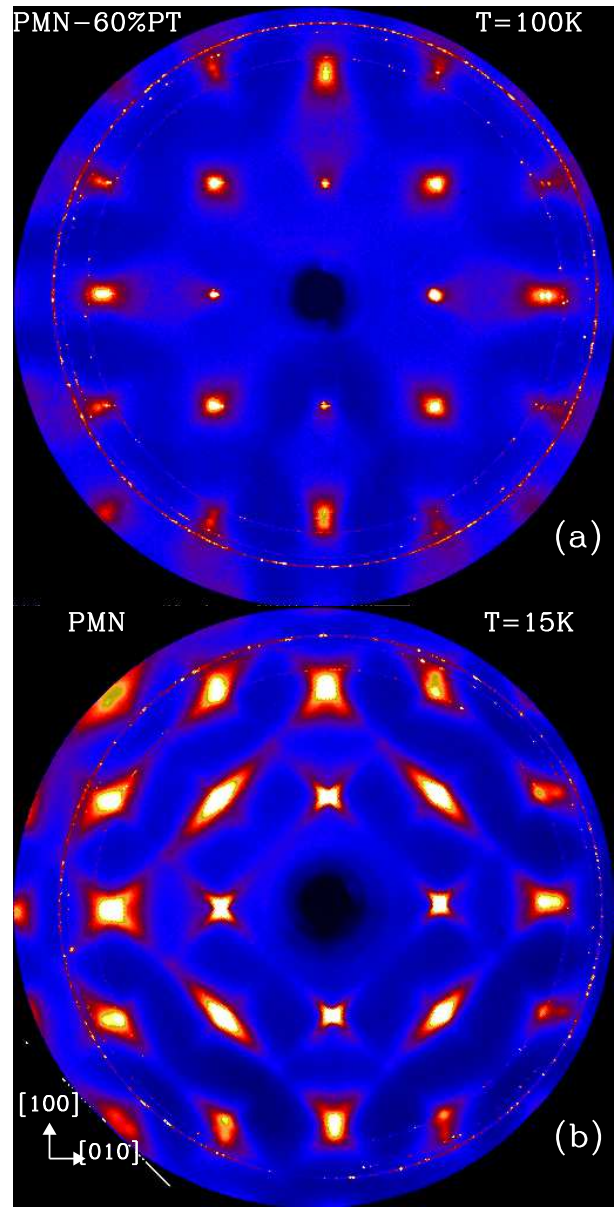}
\caption{\label{diffuse} X-ray scans done using a CCD detector at X17B.  The upper panel plots diffuse scattering in PMN-60PT and the lower panel illustrates the diffuse scattering measured in PMN.  The data have been normalized by the background.  The diffuse scattering geometry is very different in both materials.  The weak powder lines are from the sample mount and displex.}
\end{figure}

The relaxors PMN and PZN display two types of diffuse scattering.  One is highly temperature dependent around T$_{c}$, the second (denoted as the high-temperature diffuse component) is not temperature dependent and present at the highest temperatures measured.  In PMN and PZN,  the temperature dependent diffuse scattering in the (HK0) scattering plane manifests itself as rods extending along the [110] and [$\overline{1}$10] directions.  The diffuse scattering is onset above T$_{c}$, and rises steeply through T$_{c}$ until it saturates at low temperatures.  Therefore, the diffuse scattering can be interpreted in terms of short-ranged polar correlations.  At high temperatures, exceeding the Burns temperature T$_{d}$, the second type of diffuse component (high temperature diffuse) is observed along the predominately longitudinal direction (ie. parallel to $\bf{Q}$) and has been interpreted as resulting from chemical short range order.~\cite{Hiraka04:70}  The scattering from chemical short-range order is considerably weaker than the low-temperature diffuse scattering and its intensity is not strongly temperature dependent.

We conducted searches for diffuse scattering in PMN-60PT using both neutron and x-ray diffraction.  Neutron measurements were made by conducting mesh scans near the (210) and (100) Bragg positions.  X-ray measurements were made using a CCD detector such that many Bragg peaks could be simultaneously studied. Scans were done both above and below T$_{c}$ to check for any possible temperature dependence.  Both techniques gave consistent results.

\begin{figure}[t]
\includegraphics[width=8cm] {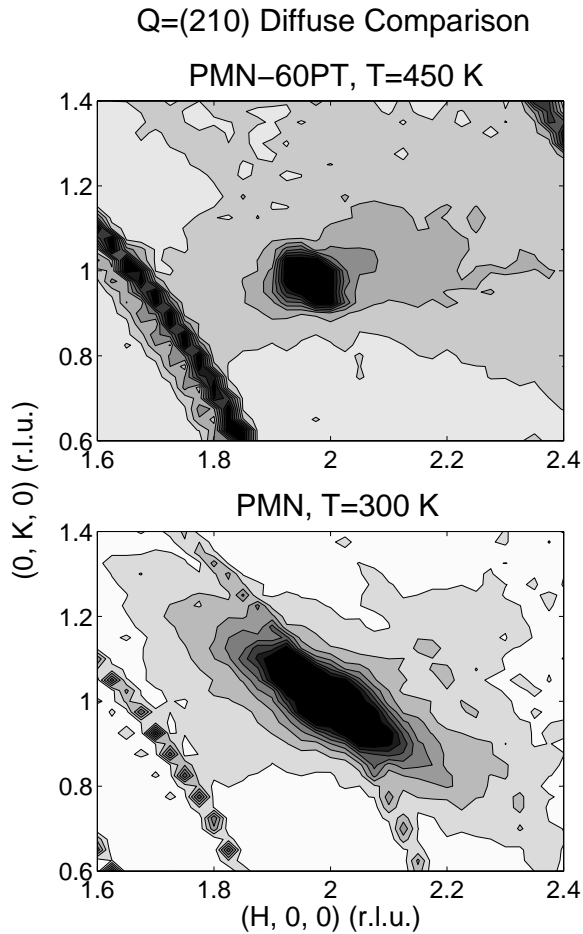}
\caption{\label{diffuse_neutron_compare} Contour scans around the (210) Bragg peak measured in PMN and PMN-60PT.  The data have been normalized to the intensity of an acoustic phonon measured at Q=(2, 0.14, 0) in both samples.  The diffuse scattering geometry is very different in both materials with an absence of the [$\overline{1}$ 1 0] diffuse scattering in PMN-60PT.  The powder ring at lower momentum transfer is due to Aluminum in the sample mount.  The powder line through the (210) point in PMN is indicative of the presence of a small amount of powder in the sample.}
\end{figure}

The main result illustrating diffuse scattering in PMN-60PT is displayed in Fig. \ref{diffuse} which plots surveys done in both PMN-60PT and PMN (for comparison purposes) using 75 keV x-rays at X17B, NSLS.  The incident beam was aligned along the [001] direction such that reflections of the form (HK0) could be studied.  All measurements presented here were done in transmission mode to eliminate any possible surface contamination. Measurements were conducted at low-temperatures where the optic mode is observed to be both underdamped and to shift to high energies (see next section), thereby minimizing effects of thermal diffuse scattering from phonons. This technique is very similar to that used elsewhere.~\cite{Xu04:70}  The data for both PMN and PMN-60PT were normalized to the same scale through a measurement of the background at the edge of the detector.  

Fig. \ref{diffuse} clearly shows that the diffuse scattering in PMN-60PT is considerably weaker than that in PMN.  PMN shows strong diffuse streaks aligned along the [$\overline{1}$10] directions, whereas PMN-60PT shows no such diffuse scattering.  A small amount of diffuse scattering along the predominately longitudinal direction is observed from the x-ray scans in PMN-60PT.  The geometry is similar to the scattering attributed to chemical short-range order in PMN.  To investigate this scattering in more detail, we used elastic neutron scattering.

To make a direct comparison of the diffuse intensity in PMN-60PT to that of PMN, we measured the diffuse intensity around (210) using mesh scans conducted at the N5 and C5 thermal triple axis spectrometers, Chalk River.  The result is presented in Fig. \ref{diffuse_neutron_compare} with both scans normalized by the intensity of a transverse acoustic phonon measured from a constant-Q scan at (2, 0.14, 0).  The phonon normalization corrects for both volume differences and spectrometer efficiency.  The comparison in Fig. \ref{diffuse_neutron_compare} shows that any diffuse scattering elongated along [$\overline{1}$ 1 0] is very weak or absent in PMN-60PT and confirms the result derived from high-energy x-rays.   However, there is a clear diffuse scattering component elongated along the longitudinal direction in PMN-60PT, which is of similar strength to that observed in PMN along the same direction.  The longitudinal diffuse scattering was first observed by Hiraka \textit{et al.} using cold neutrons.~\cite{Hiraka04:70}  The scattering was observed to be present at the highest temperatures studied and to be temperature independent.  It was concluded that the longitudinal diffuse scattering was most likely due to short-range chemical order on the B site. The observation of the weak high temperature diffuse component also sets the sensitivity of the experiment and indicates that if a strong diffuse component was present in PMN-60PT, it would have been observed in these experiments.  

\begin{figure}[t]
\includegraphics[width=8cm] {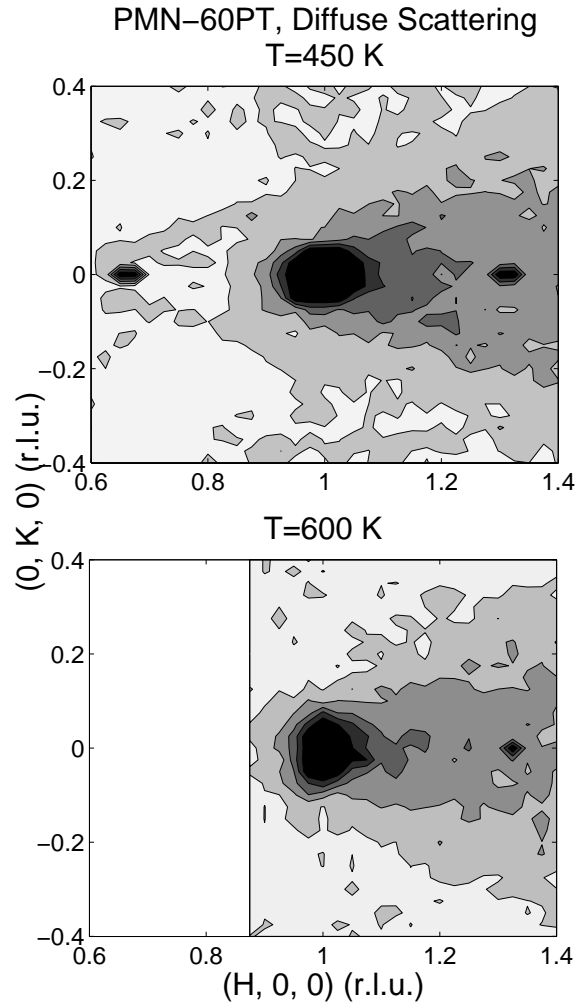}
\caption{\label{diffuse_100} Contour scans around the (100) Bragg peaks above and below T$_{c}$.  No strong temperature dependent butterfly diffuse scattering pattern (as observed in PMN) is observed in PMN.  A temperature independent bowtie pattern is observed indicating that high temperature short range chemical order is present in PMN-60PT.}
\end{figure}

Fig. \ref{diffuse_100} illustrates contour scans around the (100) Bragg peak above and below T$_{c}$.  A clear absence of any diffuse rods along the [110] and [$\overline{1}$10] is observed.  The weak scattering along the longitudinal direction is similar to the high-temperature diffuse present in pure PMN.   The weak diffuse scattering observed in PMN-60PT is temperature independent and does not change near T$_{c}$.  This contrasts to the [1$\overline{1}$0] type diffuse scattering in PMN which grows substantially near T$_{c}$.  It is also different from that in PZN, where the diffuse scattering intensity shows a peak near T$_{c}$, suggestive of a critical component which is directly related to a structural transition.  In PMN-60PT, no such behavior is observed.

The high-temperature diffuse cross section is likely due to ordering on the B site between the Mg and Nb ions as suggested previously for pure PMN.  The scattering intensity in both the (210) and (100) zones of the weak diffuse component in PMN-60PT is clearly more intense on the larger $\bf{Q}$ side than the low-$\bf{Q}$ side.  This is consistent with the diffuse cross section resulting from atomic displacements where $I \sim |Q \cdot \epsilon|^2$, where $\epsilon$ is the atomic displacements.  Theoretical studies of the ordering on the Mg and Nb sites have suggested that short-range order would lead to displacements on the Pb site.~\cite{Burton99:60} The ordering is expected to occur at temperatures in excess of $\sim$ 1000 K, consistent with the lack of any temperature dependence below 800 K measured in this experiment.  This conclusion is supported by NMR data which show a small amount of ordering on the B site at high temperatures as evidenced by two components to the $^{93}$Nb lineshape.~\cite{Laguta03:67} High resolution electron microscopy and polarized Raman studies have also provided strong direct evidence for short-range ordering on the B site.~\cite{Boulesteix94:108,Svitelskiy03:68} It is interesting to note that near (100), the diffuse scattering peaks at around $1/3$ where in pure PMN it is observed to peak at around the incommensurate position of $\sim$ 0.1.  The peak at around $1/3$ is expected in a fully ordered Pb(B$'$$_{1/3}$B$'$$'$$_{2/3}$)O$_{3}$ with a B$'$-B$'$$'$-B$'$$'$-B$'$ type structure providing evidence that the high-temperature diffuse scattering is associated with ordering on the B site. Therefore, the addition of Ti through doping with PbTiO$_{3}$ may promote ordering on the B site and possibly reduce the structure disorder.

In the relaxor ferroelectrics PMN and PZN, the $\langle$110$\rangle$ rod type temperature dependent diffuse scattering comes from the formation of polar nanoregions, or regions of ferroelectric order in a paraelectric background.  The absence of diffuse scattering in PMN-60PT suggests that the polar nanoregions are replaced by a long-range ordered tetragonal structure.  The presence of weak diffuse scattering along the longitudinal direction suggests that short-range chemical order still exists in highly doped PT.  The short-range chemical order closely resembles the B$'$-B$'$$'$-B$'$$'$ type ordering expected based on stoichiometry.  The absence of $\langle$110$\rangle$  rod type temperature dependent diffuse scattering in PMN-60PT marks a clear difference between PMN-60PT, and the prototypical relaxors PMN and PZN.  

\section{Phonons}

In classic ferroelectrics, the phonons play an important role in the formation of a long-ranged polar phase as a transverse optic phonon softens in energy to zero frequency at the critical temperature.  PbTiO$_{3}$ displays classic dynamic behavior near T$_{c}$ with the zone-centre frequency approaching zero frequency at the critical temperature.  Relaxors, however, show a broad optic mode near the zone-centre and do not completely soften.  In understanding the structural transition in PMN-60PT, it is important to characterize the dynamic behavior and compare it to both PMN and pure PbTiO$_{3}$.  

\subsection{Lineshape and Data Analysis}

\begin{figure}[t]
\includegraphics[width=8cm] {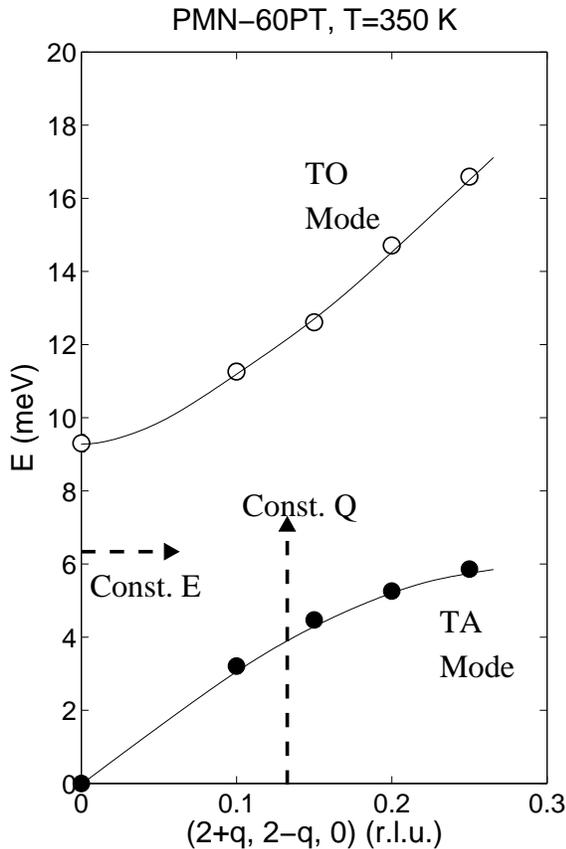}
\caption{\label{dispersion} The dispersion of the TA$_{2}$ and TO$_{2}$ is plotted near the zone centre at Q=(220) for T=350 K.  For higher temperatures, near T$_{c}$, the zone centre optic frequency was extracted by extrapolating to the zone centre from non-zero-q.  The dashed arrows indicate the two types of scans (constant E and Q) used to investigate the phonons.}
\end{figure}

The phonons of most interest in ferroelectrics are the low-energy transverse acoustic and optic modes.  We investigated the dispersion and lineshape of the phonon modes from both constant-Q (where energy is scanned fixing the momentum transfer) and constant energy (where the momentum transfer is held fixed while scanning energy) scans.  The difference between the two methods is illustrated in Fig. \ref{dispersion}.  The dispersion of the TA and TO modes is illustrated in Fig. \ref{dispersion} as measured from constant Q scans near the (220) position at T=350 K.  The temperature dependence of the optic mode frequency is of the greatest interest as it can be directly related to the dielectric constant.

We have measured both the T$_{1}$ and T$_{2}$ phonons by conducting constant-Q scans in the (200) and (220) zones.  In the (200) zone, we have studied transverse acoustic T$_{1}$ phonons with the propagation vector along the [010] direction and polarization along [100].  For scans near (220), we have investigated T$_{2}$ phonons with propagation vector along [1$\overline{1}$0] and polarization along [110].  Therefore, any anisotropy in the lattice dynamics can be studied.  

To extract parameters such as the frequency position and linewidth as a function of temperature, a particular lineshape must be convolved with the resolution function and fit to the data.  The measured intensity from a triple-axis spectrometer is directly proportional to S($\bf{q}$, $\omega$), which is related to the imaginary part of the susceptibility through the fluctuation dissipation theorem~\cite{Shirane:book},

\begin{eqnarray}
\label{sqw} S({\bf{q},}\omega) = {1\over \pi} [n(\omega)+1] \chi''({\bf{q}}, \omega).
\end{eqnarray}

\noindent We have used the following form for the imaginary part of the susceptibility given by the antisymmetrized linear combination of two Lorentzians,

\begin{eqnarray}
\label{SHO} \chi''({\bf{q},}\omega)={{A\times\Gamma}\over {[\Gamma(\omega)^{2}+
\left(\hbar \omega- \hbar \omega_{0}(\bf{q})\right)^{2}}]} \\
- {{A\times\Gamma}\over {[\Gamma(\omega)^{2}+ \left(\hbar \omega+\hbar
\omega_{0}(\bf{q})\right)^{2}}]},
\nonumber
\end{eqnarray}

\noindent where $\Gamma(\omega)$ is the frequency dependent half-width-at-half-maximum (HWHM), $\omega_{0}(\bf{q})$ is the undamped phonon frequency, and A is the amplitude. For acoustic modes, we have approximated the dispersion to be linear $\omega_{0}$=c$|{\bf{q}}|$.  For optic modes, we have set the dispersion to have the form described by,

\begin{eqnarray}
\label{optic} \omega^2(T)=\Omega_{0}^2(T)+\alpha q^2, 
\end{eqnarray} 

\noindent where $\hbar\Omega_{0}(T)$ is the temperature \textit{dependent} zone centre frequency and $\alpha$ is the temperature \textit{independent} slope.  This is the same lineshape used previously for PMN and PZN.~\cite{Stock04:69}

\begin{figure}[t]
\includegraphics[width=7cm] {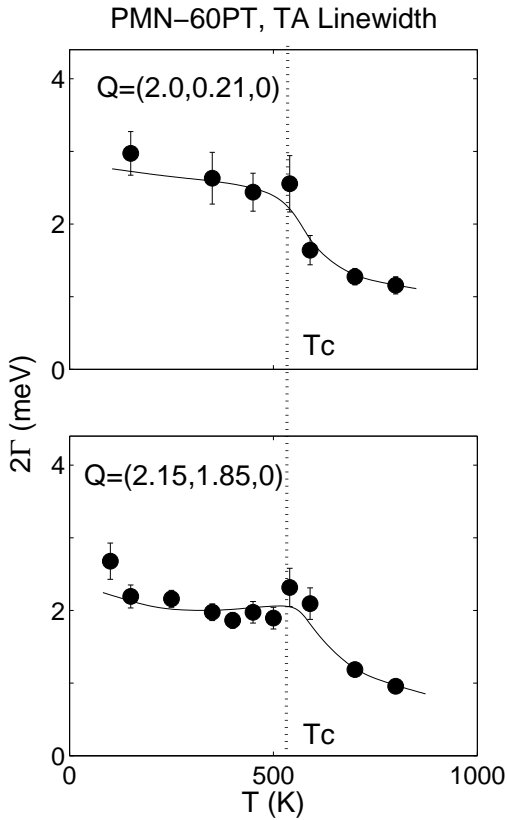}
\caption{\label{TA_linewidth} The transverse acoustic linewidth is plotted as a function of temeprature for both the T$_{1}$ (near (200)) and T$_{2}$ (near (220)) modes.  A significant broadening is observed near T$_{c}$.  This is interpreted as resulting from domains.}
\end{figure}

\subsection{Transverse Acoustic: Comparison with PMN}

\begin{figure}[t]
\includegraphics[width=7cm] {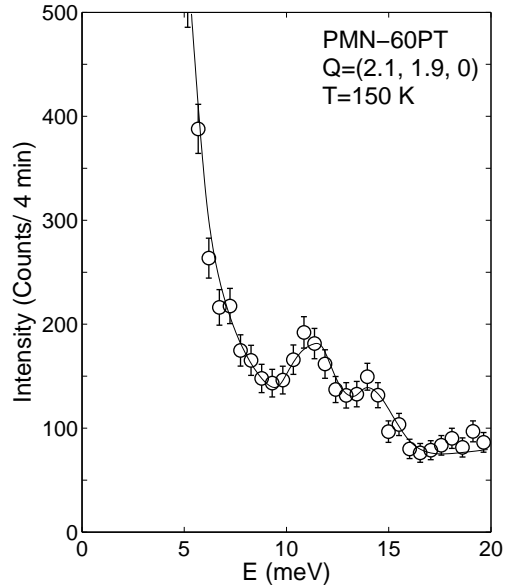}
\caption{\label{tetra_domains} Constant Q scan at Q=(2.1, 1.9, 0) taken at T=150 K, below the transition from cubic to tetragonal transition.  The optic mode is clearly seen to be split into two peaks indicative of the underlying tetragonal domain structure in the sample.}
\end{figure}

The acoustic mode in PMN was been studied in detail by several groups.~\cite{Nab99:11,Wakimoto02:65}  A subtle broadening of the acoustic mode was observed in PMN over a broad range in temperature extending from the critical temperature T$_{c}$ to the high temperature Burns temperature T$_{d}$.  The broadening in the TA mode was very small compared with the significant damping of the TO mode which was observed to occur over the same temperature region.  In PMN, the broadening of the TA peak has been associated with a dampening from polar domains formed near the Burns temperature.  Given the absence of any strong diffuse scattering in PMN-60PT, it is important to study the linewidth of the acoustic mode to see if any broadening above T$_{c}$ is observed.

We have investigated the temperature dependence of both the TA$_{1}$ and TA$_{2}$ modes scanned near the (200) and (220) zones respectively.  The results at Q=(2, 0.21, 0) and Q=(2.15, 1.85, 0) are plotted in Fig. \ref{TA_linewidth}.  To extract the full-width from constant Q scans, we fit the equation given by the antisymmetrized linear combination of two Lorentzians stated previously.   Unlike PMN which shows a significant sharpening of the TA mode near T$_{c}$, PMN-60PT displays a significant broadening of both the T$_{1}$ and T$_{2}$ modes below T$_{c}$.  We attribute this broadening to formation of domains in the tetragonal phase.  

To verify the presence of domains in the bulk and hence different acoustic and optic slopes near the zone center, we searched for a splitting of the acoustic or optic peaks at low-temperatures, well away from the structural transition where the difference in elastic constants is the largest.  A splitting of the TO mode was observed at low temperatures (as illustrated in Fig. \ref{tetra_domains}) near the (220) Bragg peak.  This splitting confirms the presence of domains and hence a tetragonal distortion.  The observation also confirms that the tetragonal distortion is a \textit{bulk} effect and not due to a near surface region as observed in PMN and PZN.

The temperature dependence of the TA linewidth in PMN-60PT is very different from that in PMN.  PMN-60PT shows no clear broadening of the TA mode above T$_{c}$.  This is consistent with the fact that the diffuse scattering is very weak in PMN-60PT and also that the broadening originates from the polar domains in pure PMN.  Therefore, our results on PMN-60PT directly associate the broadening of the TA peak in PMN with the presence of polar domains.  The result also supports our assertion that the polar regions in PMN-60PT are either small or absent.

\subsection{Overdamped Zone Centre Transverse Optic Mode}

\begin{figure}[t]
\includegraphics[width=9cm] {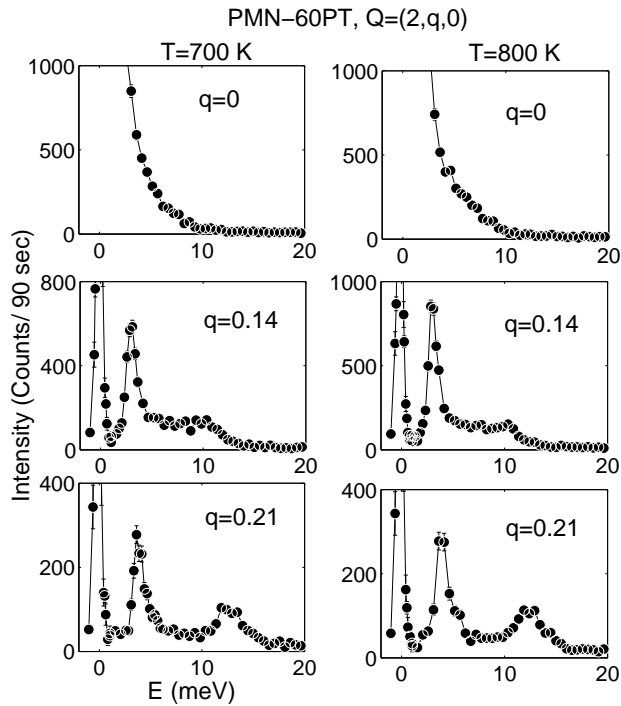}
\caption{\label{constQ_200} Typical constant-Q scans near the (200) position at 700 K and 800 K illustrating the highly damped TO phonon near the zone centre.  The TO mode recovers at larger q=0.21, indicating that the anomalous dampening is restricted to the zone centre. The data was taken at the C5 spectrometer.}
\end{figure}

\begin{figure}[t]
\includegraphics[width=9cm] {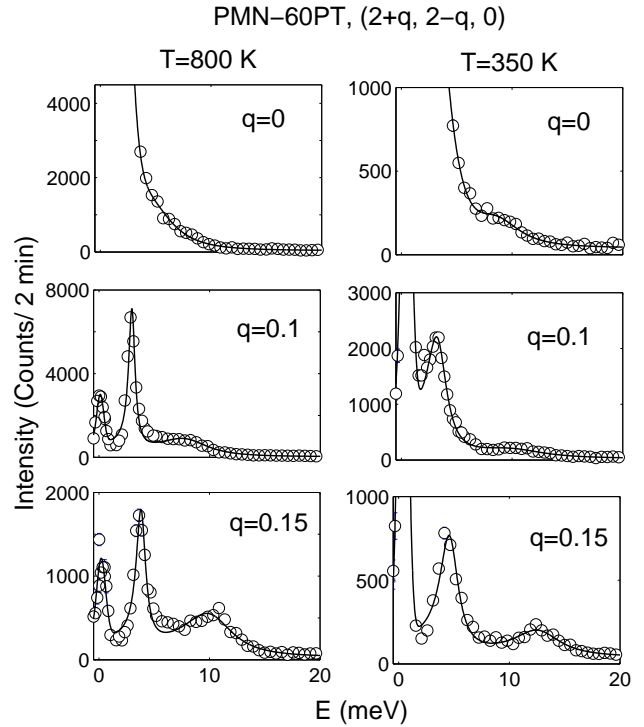}
\caption{\label{constQ} Typical constant-Q scans near the (220) position at 350 K and 800 K used to extrapolate to the zone centre.  The data were taken at the N5 spectrometer. At high-temperatures, the TO mode is broad and not observable at q=0 with a constant-Q scan.  The data at 350 K indicate a recovery of the TO mode, especially clear at non-zero q=0.15.}
\end{figure}

\begin{figure}[t]
\includegraphics[width=7cm] {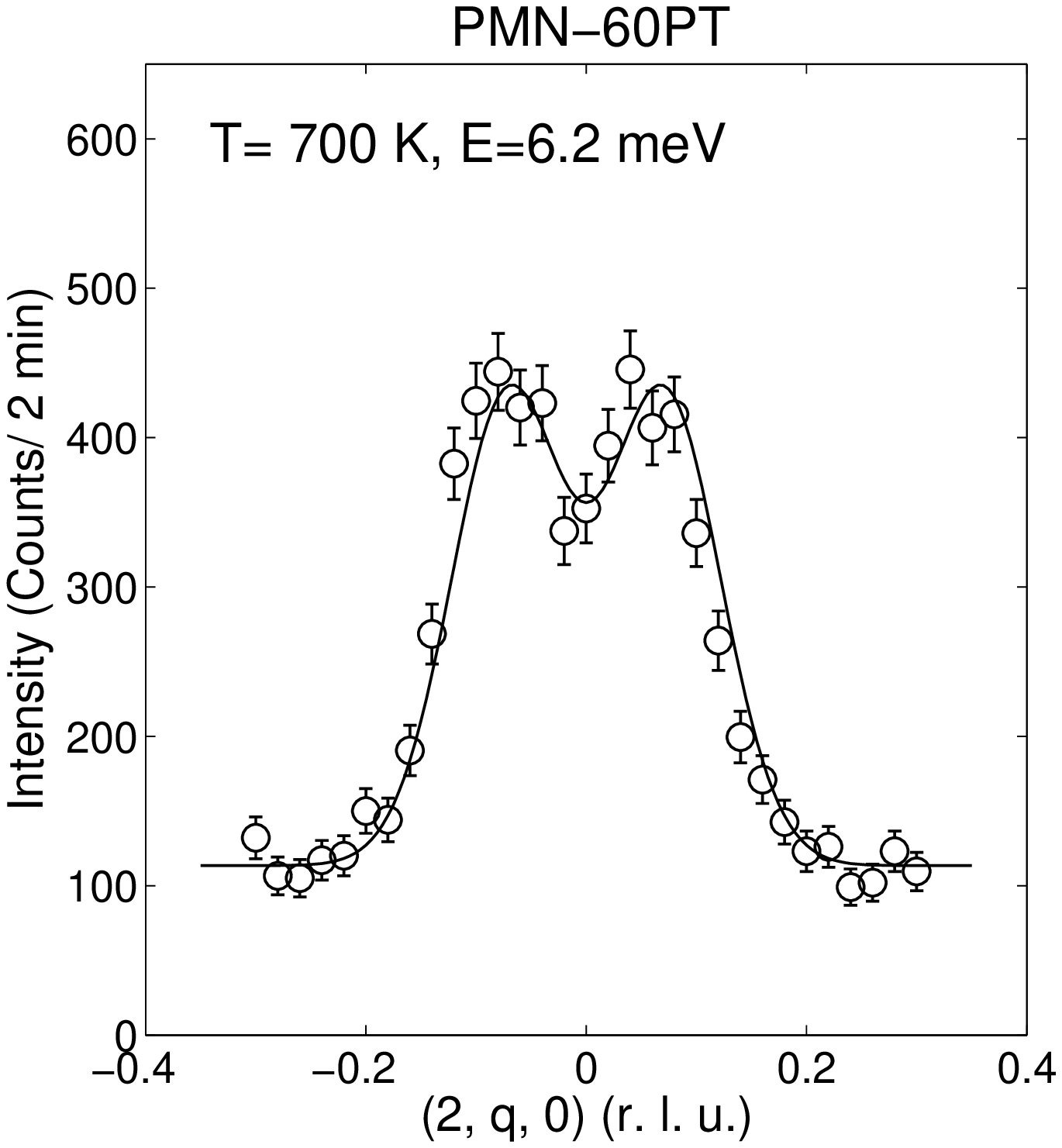}
\caption{\label{constE} A constant energy scan cutting through the two branches of the optic mode at 700 K, E=6.2 meV is plotted.  Even though the optic mode does not show a sharp peak in a constant Q scan, it is clearly present from a constant energy (const. E) scan.  This indicates that the mode is strongly damped near T$_{c}$.}
\end{figure}

As stated previously, the optic mode plays an important role in the structural distortion in ferroelectrics.  The optic mode in PMN-60PT was investigated by conducting both constant-Q and constant energy scans near the (220) and (200) positions.  Typical scans are illustrated in Figs. \ref{constQ_200} and \ref{constQ}.  At the highest temperatures investigated, no propagating transverse optic mode is observed at the zone center (q=0).  Scans at small but non-zero $q$ in both the (200) (Fig. \ref{constQ_200}) and (220) (Fig. \ref{constQ}) zones at 800 K show the optic mode to be broad and to recover at only non-zero $|q|$.  At lower temperatures, below Tc, a recovery of the optic mode is measured as a TO peak is observed at the zone centre.  It is interesting to note that no well defined optic mode was observed near the zone centre between Tc and 800 K indicating that optic mode remains damped with little recovery to very high temperatures.  The presence of an overdamped optic mode is further verified by a constant energy scan through the two branches of the optic mode.  Fig. \ref{constE} plots a scan along the (2,q,0) direction at a fixed energy transfer of 6.2 meV.  The scan cuts through the two dispersive branches of the optic mode giving rise to two peaks symmetrically displaced from q=0.  The fact that no well defined peak is observed in a constant-Q scan, but observed in a constant energy scan, indicates that the TO mode is heavily damped and has a short lifetime associated with it.

The presence of a highly damped TO mode over a broad range in temperature near T$_{c}$ is very similar to the behavior observed in PMN.  In PMN, the optic mode is highly damped between T$_{c}$ ($\sim$ 200 K) and the Burns temperature T$_{d}$ ($\sim$ 600 K).  In PMN-60PT, the optic mode is damped at least over the range of temperatures of $\sim$ 500 K - 800 K.  Assuming that the temperature ranges scale with T$_{c}$, as they do in PMN and PZN, the broad TO mode over such a large temperature range is consistent with the  behavior observed in PMN.

The presence of a broad TO mode similar in lineshape and temperature dependence to that of PMN, is very surprising as diffuse measurements show that there is an absence of any diffuse scattering indicative of polar nanoregions.  In PMN, one possible model for the broad TO mode was that polar nanoregions, which are formed at the Burns temperature, significantly damp the optic mode.  This conclusion was drawn from the observation that the diffuse scattering and the dampening were onset at nearly the same temperature.  Our measurements on PMN-60PT show that polar nanoregions are absent and are replaced by a long-range ordered structure, indicating that such an interpretation for the dampening of the soft mode needs to be revaluated.   

Even though the polar nanoregions are small in PMN-60PT and are not directly observable from diffuse scattering, we do observe diffuse scattering along the longitudinal direction indicative of short-range chemical order signifying the presence of a substantial amount of structural disorder.  The B site in this PbBO$_{3}$ structure consists of a mixture of B=Mg$^{2+}$, Nb$^{5+}$, and Ti$^{4+}$. It is likely that the large difference in oxidation states on the B site will introduce large electric field gradients thereby introducing significant random fields.   We therefore conclude that random fields introduced from the substantial amount of chemical disorder is the cause for the overdamped TO mode rather than any disorder introduced through polar nanoregions.  The high temperature diffuse scattering represents some form of domains or localized order in the sample, possibly from chemical short-range order as suggested by Hiraka \textit{et al.}~\cite{Hiraka04:70}  A comparison to other disordered structures is given later in the paper.

\begin{figure}[t]
\includegraphics[width=8cm] {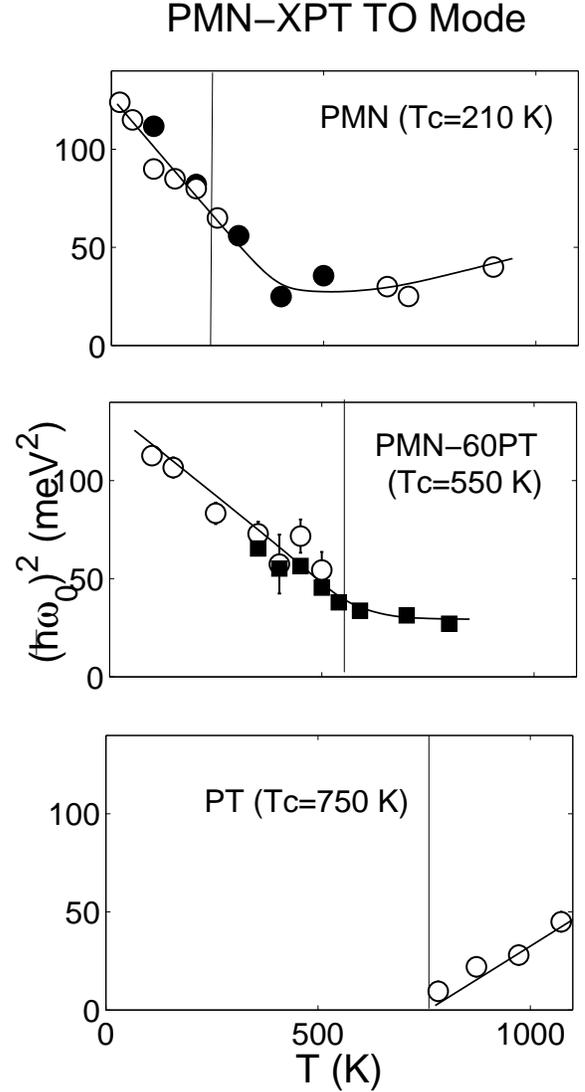}
\caption{\label{soft_mode} A comparison of the soft mode temperature dependence in PMN, PMN-60PT, and pure PT.  The data for PT were taken from Shirane \textit{et al.}~\cite{Shirane70:2}  The filled symbols represent values obtained by extrapolating from non-zero q to the zone centre.  The open circles are TO energy values obtained from a direct cosntant-Q scan at the zone center.}
\end{figure}

A comparison of the soft-mode temperature dependence in PMN, and PMN-60PT, and pure PT is presented in Fig. \ref{soft_mode}.  The open circles for PMN and PMN-60PT were taken from constant-Q scans at the zone centre.  For temperatures near T$_{c}$, the zone centre frequency is very difficult to determine as the linewidth becomes large.  We therefore conducted constant-Q scans at non-zero $q$ away from the zone center and then extrapolated to the zone center using the formula $\omega^2=\Omega_{0}^2+\alpha q^2$.  $\Omega_{0}$ is the zone centre frequency and $\alpha$ was fixed to be temperature independent.  Based on this method, the zone centre frequency can be extracted even though the TO mode is overdamped near the zone centre and is represented by the filled points in Fig. \ref{soft_mode}.

The soft mode behaviors of PMN and PMN-60PT around T$_{c}$ are very similar.  In both compounds the TO does not completely soften to zero energy, however decreases in energy to $\sim$ 5 meV.  Both compounds show a clear recovery of the TO mode below a characteristic temperature.  It is interesting to note that in PMN-60PT, the recovery occurs below T$_{c}$ where a well defined structural transition from a cubic unit cell to tetragonal cell occurs and where there exists a peak in the dielectric response. However, in PMN, the recovery starts at a much higher temperature of about 400 K.  The high temperature recovery maybe more closely associated with the peak in the dielectric response, rather than the temperature where the structural transition is onset under the application of an electric field.  We emphasize that in pure PMN, T$_{c}$ is defined as the temperature at which a sharp anomaly appears in the dielectric susceptibility under the application of an electric field.  In zero field, there is no detectable structural transition in the bulk of PMN.

The temperature dependence of the soft transverse-optic mode in PMN and PMN-60PT is very different from that in PbTiO$_{3}$.  The slope of the soft-mode above T$_{c}$ is dramatically altered from the pure material and is not described simply by mean-field theory which predicts the slope of the temperature dependence below T$_{c}$ to be twice that above T$_{c}$.  As noted by Halperin and Varma~\cite{Halperin76:14}, in the presence of defects the temperature dependence of the soft-mode can be dramatically altered near T$_{c}$.  Therefore, at least qualitatively, the temperature dependence of the soft-mode can be interpreted in terms of slowly relaxing or frozen defects. 

In PMN, there are two well defined temperature scales defined by the behavior of the soft transverse optic mode.  There is a high temperature region where the optic mode reaches a minimum in energy and begins to recover.  There is also a second temperature scale at T$_{c}$ where an anomaly appears in the dielectric susceptibility under the application of an electric field.  In contrast to this, PMN-60PT displays one temperature scale where the optic mode recovers in energy and a structural transition occurs.  PMN-60PT does display a broad TO mode analogous to PMN.   In contrast to both PMN and PMN-60PT, pure PbTiO$_{3}$ displays a well defined soft TO mode which goes to zero, initiating a structural transition.  Therefore, in terms of optic phonons, PMN-60PT is very similar to the behavior of prototypical relaxor ferroelectrics like PMN and PZN.  The temperature dependence and lineshape of the TO mode is, however, very different from the behavior in PbTiO$_{3}$.

\section{Conclusions and Discussion}

We have shown that there is no strong temperature dependent diffuse scattering in PMN-60PT, in contrast to pure PMN.  This has been demonstrated using survey scans done with penetrating 75 keV x-rays, neutron elastic scattering, and also through the absence of any anomalous broadening in the TA mode above T$_{c}$.  The absence of diffuse scattering in PMN-60PT implies that that polar nanoregions are now replaced by large domains and long-range structural order.  We have also shown that there exists a broad transverse optic mode near the zone centre which is very similar to pure PMN. Therefore, the broadening of the TO mode (and hence the waterfall effect) cannot arise from the formation of polar nanoregions.  These results show that the elastic scattering is very similar to a classic ferroelectric, PbTiO$_{3}$, however, the dynamics including the soft optic mode are nearly identical to a relaxor ferroelectric with no well defined structural transition.

To investigate a possible microscopic model to consistently explain the diffuse and optic phonon data in PMN and PMN-60PT, we consider the origin of the two types of diffuse scattering in more detail.  It has been shown by Hirota \textit{et al.}~\cite{Hirota02:65} and Vakhrushev \textit{et al.}~\cite{Vak89:90} that the atomic displacements associated with the diffuse scattering consists of a centre of mass conserving component, associated with the condensation of the optic mode, and a centre of mass shift.  A recent detailed analysis of the acoustic and optic coupling in PMN has shown that mode coupling between the acoustic and optic modes is weak and therefore it is not likely that the centre of mass shift occurs at the same temperature at which the optic mode softens.~\cite{Stock04:xx}  Another explanation is that the centre of mass shift is present at much higher temperatures and may be associated with the short-range chemical order indicated by the presence of high-temperature diffuse scattering.  In this scenario, centre of mass shifted domains occur at a high temperature and give rise to diffuse scattering along the longitudinal direction.  These domains introduce defects into the crystal lattice and significantly damp the optic mode as it softens in energy near T$_{c}$.  The presence of a centre of mass shift at high temperatures therefore consistently explains the presence of both the high-temperature diffuse scattering as well as the significant damping of the optic mode near T$_{c}$ throughout the PMN-xPT phase diagram.  The origin of this disorder maybe associated with the disorder on the B site, as the longitudinal diffuse scattering disappears and the damping of the transverse optic mode disappears in the limit of pure PbTiO$_{3}$.

The presence of locally shifted regions in the crystal will ultimately introduce defects into the crystal lattice.  One of the simplest examples of a crystal with defects is the H$_{2}$-D$_{2}$ mixed crystal.~\cite{Powell75:12}  Powell and Nielsen describe a theory of phonons in a crystal with defects in which the cross section has two components giving rise to coherent scattering.  The first is a modified phonon cross section resulting in well defined peaks in energy resulting from the excitation of phonons.  The second is a broad resonance diffuse term with a width which is a measure of the density of host modes into which the impurity mode can decay.  The diffuse component was difficult to observe on its own due to the fact that it was extremely broad in momentum and energy and the cross-section depended on the difference of the mixed atoms.  It is interesting to note that that the phonons in H$_{2}$-D$_{2}$ have some strong similarities to those in the mixed ion relaxor systems.  In particular, the low energy TO mode was observed to significantly broaden in energy at a characteristic wavevector where the TO dispersion overlaps with the broad resonance component in the defect phonon cross section.  At that particular energy, the TO mode has a short lifetime due to an increase in the number of decay channels.  The TO mode was damped considerably despite the fact that TA mode does not show any significant effect.  This is nearly identical to the behavior in the relaxor ferroelectrics where there is also a characteristic wavevector where the TO mode becomes significantly damped.~\cite{Gehring01:63}  We emphasize that this resonance is not a well-defined mode, but a continuum of states providing a new decay channel when a propogating mode (like the TO mode) crosses the correct energy scale.

The presence of a density of states or a broad resonance contribution to the scattering is further supported by Raman and dielectric measurements.  Previous dielectric experiments on PMN~\cite{Kambaxx:04} have shown the presence of a finite density of states at low-energies.  This appeared in addition to the soft-mode response which agreed with the soft transverse-mode measured with neutrons.  The calculated neutron scattering spectra at low temperatures displayed the presence of two peaks, the higher energy peak is the soft-mode measured at the zone-centre with neutrons, while the lower-energy peak has not be observed with neutrons.  A similar low-energy peak in the dielectric response has also been observed in PST.~\cite{Kambaxx:05}  The new lower energy peak, might represent a broad resonance suggested here which results in significant dampening of the optic mode over a certain range in energy and momentum.  

The unusual temperature dependence of the TO linewidth in PMN and PMN-60PT can be interpreted in terms of this defect model.  Both compounds show a broad TO mode which becomes overdamped at a characteristic energy, and hence wavevector.  The fact that broadening only occurs over a finite temperature region can be explained by the fact that the TO mode softens considerably with temperature, and therefore only enters the correct characteristic energy region over a finite temperature scale.   The diffuse or broad mode in PMN and PMN-60PT can be associated with defect states introduced by the mixed valency.  These low energy states provide extra decay channels for the TO mode over a finite energy range.  This model is consistent and is expected based on a random field model introduced recently to describe the universal properties observed in PMN and PZN.~\cite{Stock04:69}

The idea of defects causing extra channels at low-energies for the TO mode to decay into is also consistent with a recent model proposed to describe the dynamics by Yamada and Takakura.~\cite{Yamadaxx:9573}  They proposed that the broadening of the TO mode was due to coupling to a random pseudospin variable.  Yamada and Takakura suggested that a microscopic origin for this could be the random hopping of the Pb$^{2+}$ ion as it has been found that the equilibrium position for Pb$^{2+}$ is slightly displaced from the high symmetry position.  Such a model has been supported by $^{207}$Pb NMR experimental data.~\cite{Blinc03:91} Yamada and Takakura described the phonons in PMN utilizing the Langenvin equation which describes the atomic motion under a random force.  The phonon model explained the broadening of the TO mode near the zone center in accordance with experiment.  This model is identical to that proposed by Powell and Nielsen to explain defects in H$_{2}$-D$_{2}$.  Yamada and Takakura consider the Pb$^{2+}$ ions to be hopping between sites at a characteristic frequency.  When the soft-optic mode energy is close to the hopping frequency, significant dampening will occur of the extra channels by which the phonon can decay into.

It is important to note that PMN displays two very distinct temperature scales, a high temperature scale where the diffuse scattering is onset (T$_{d}$) and a lower temperature (T$_{c}$) where a structural transition occurs under the application of an electric field.  These two temperature scales have been interpreted consistently using a random field model.~\cite{Stock04:69}  Random fields have also been suggested by other groups to explain the dynamics and static properties of the relaxors.~\cite{Westphal92:68,Pirc99:60,Fisch03:67}  In the random field model proposed, the temperature scales were suggested to result from two universality classes.  At high temperatures, the temperature energy scale can be thought to be much higher than any cubic anisotropy in the system effectively making the system Heisenberg like, or possessing a continuous symmetry.  At lower temperatures, the cubic anisotropy becomes important and the system becomes more Ising-like with a discrete symmetry.  Both PMN and PZN show two clear temperature scales.  In model magnetic systems it was found that systems with a continuous symmetry are very sensitive to the presence of random fields.  In contrast, systems with a discrete (or Ising) symmetry are robust to the presence of random fields.

In PMN-60PT, there is distinct absence of two temperature scales.  Both a structural transition and a recovery of the transverse optic mode occur at the sample temperature.  Any model describing the PMN-XPT phase diagram must be able to consistently describe the behavior in PMN, PMN-60PT, and PbTiO$_{3}$.  We propose that for increased PT doping the cubic anisotropy becomes more important, as well, the random field must ultimately decrease in the extreme case of pure PbTiO$_{3}$, with no mixed valency.   In the case of PMN-60PT, the cubic anisotropy must become important as the system undergoes a structural transition to a tetragonal phase, with the atoms shifted along the [100] directions.  Therefore, the system can be thought of as effectively having a discrete symmetry and properties similar to model magnetic systems in the presence of random fields.~\cite{Feng96:55}

The presence of a stronger cubic anisotropy in PMN-60PT in comparison to PMN is further reflected in the slopes of the TO$_{1}$ and TO$_{2}$ modes.  In PMN, the slopes (as defined by $\hbar^{2}\alpha$ in Eqn. \ref{optic}) are nearly identical with $\hbar^{2}\alpha_{T1}$=590 meV$^{2}$ \AA$^{2}$ and $\hbar^{2}\alpha_{T2}$=460 meV$^{2}$ \AA$^{2}$.  In PMN-60PT, the difference is much more dramatic with $\hbar^{2}\alpha_{T1}$=955 meV$^{2}$ \AA$^{2}$ and $\hbar^{2}\alpha_{T2}$=685 meV$^{2}$ \AA$^{2}$ indicating that the dynamics are more anisotropic in PMN-60PT than in PMN.  Our measurements show that the PMN-$x$PT phase diagram may provide a unique example of random fields in structural transitions and an opportunity to study the dynamics under a random field where the universality class can be continuously tuned from a continuous to a discrete symmetry.  Clearly, further measurements are required for intermediate PT concentrations to provide further credence to this model.

We have presented a detailed elastic and inelastic scattering study of PMN-60PT.  Through the use of high-energy x-rays and elastic neutron scattering, we have shown that any strong diffuse scattering from polar nanoregions is very weak or absent in PMN-60PT in favour of a long-range structural ordered ground state with a tetragonal unit cell.  We have found weak predominately longitudinal diffuse scattering indicative of short-range chemical order.  The optic mode is overdamped around T$_{c}$ in a similar manner to that in PMN.  We have reconciled this by suggesting that the short-range chemical order provides extra channels at low-energies into which the the TO mode can decay into, analogous to the case of the defect crystal H$_{2}$-D$_{2}$ and to the Pb$^{2+}$ ion hopping model proposed by Yamada and Takakura.  We have also interpreted the presence of a well defined structural transition in the presence of defects and a broad TO mode in terms of a random field model previously proposed to explain the similarity between PMN and PZN.

\begin{acknowledgements}

We are grateful to M. Potter, L. McEwan, R. Sammon, T. Whan, and M. Watson for technical assistance.  The work at the University of Toronto was supported by the Natural Science and Engineering Research Council of Canada and the National Research Council of Canada.  We also acknowledge financial support from the U.S. DOE under contract No. DE-AC02-98CH10886, and the Office of Naval Research under Grant No. N00014-99-1-0738.   

\end{acknowledgements}



\thebibliography{}

\bibitem{Ye98:81} Z.-G. Ye, \textit{Key Engineering Materials Vols. 155-156}, 81 (1998).

\bibitem{Park97:82} S.-E. Park and T.R Shrout, J. Appl. Phys. {\bf{82}}, 1804 (1997).

\bibitem{Xu03:67} G. Xu, Z. Zhong, Y. Bing, Z.-G. Ye, C. Stock, and G. Shirane, Phys. Rev. B {\bf{67}}, 104102 (2003).

\bibitem{Stock04:69} C. Stock, R.J. Birgeneau, S. Wakimoto, J.S. Gardner, W. Chen, Z.-G. Ye, and G. Shirane, Phys. Rev. B {\bf{69}}, 094104 (2004).

\bibitem{Burns83:48} G. Burns and F.H. Dacol, Solid State Commun. {\bf{48}}, 853 (1983).

\bibitem{Jeong05:94} I.-K. Jeong, T.W. Darling, J.K. Lee, Th. Proffen, R.H. Heffner, J.S. Park, H.S. Hong, and W. Dmowski, Phys. Rev. Lett. {\bf{94}}, 147602 (2005).

\bibitem{Blinc03:91} R. Blinc, V. Lagutam and B. Zalar, Phys. Rev. Lett. {\bf{91}}, 247601 (2003).

\bibitem{Nab99:11} A. Naberezhnov, S. Vakhrushev, B. Doner, D. Strauch, and H. Moudden, Eur. Phys. J. B {\bf{11}}, 13 (1999).

\bibitem{Hlinka03:91} J. Hlinka, S. Kamba, J. Petzelt, J. Kulda, C.A. Randall, S. J. Zhang, Phys. Rev. Lett. {\bf{91}}, 107602 (2003).

\bibitem{Vak02:xx} S.B. Vakhrushev and S.M. Shapiro, Phys. Rev. B. {\bf{66}}, 214101 (2002).

\bibitem{Gehring01:63} P.M. Gehring, S.-E. Park, and G. Shirane, Phys. Rev. B. {\bf{63}}, 224109 (2001).

\bibitem{Cowley05:5584} S.N. Gvasaliya, B. Roessli, R.A. Cowley, P. Hubert, S.G. Lushnikov, cond-mat/0505584.

\bibitem{Wakimoto02:65} S. Wakimoto, C. Stock, R.J. Birgeneau, Z.G-. Ye, W. Chen, W.J.L. Buyers, P.M. Gehring, and G. Shirane, Phys. Rev. B {\bf{65}}, 172105 (2002).

\bibitem{Noheda02:66} B. Noheda, D.E. Cox, G. Shirane, J. Gao, and Z.-G. Ye, Phys. Rev. B {\bf{65}}, 224101 (2002).

\bibitem{Ye03:67} Z.-G. Ye, Y. Bing, J. Cao, A.A. Bokov, P. Stephens, B. Noheda, and G. Shirane, Phys. Rev. B {\bf{67}}, 104104 (2003).

\bibitem{Lente04:xx} M.H. Lente, A.L. Zanin, E.R.M. Andreeta, I.A. Santos, D. Garcia, and J.A. Eiras unpublished (cond-mat/0407042).

\bibitem{Feng04:16} Z. Feng, Z. Zhao, and H. Luo, J. Phys.: Condens. Matter {\bf{16}}, 6771 (2004).

\bibitem{Ohwa01:70} H. Ohwa, M. Iwata, H. Orihara, N. Ysauda, and Y. Ishibashi, J. Phys. Soc. Jpn. {\bf{70}}, 3149 (2001).

\bibitem{Shirane70:2} G. Shirane, J.D. Axe, J. Harada, and J.P. Remeika, Phys. Rev. B {\bf{2}}, 155 (1970).

\bibitem{Westphal92:68} V.Westphal, W. Kleemann, M.D. Glinchuk, Phys. Rev. Lett. {\bf{68}}, 847 (1992).

\bibitem{Pirc99:60} P. Pirc and R. Blinc, Phys. Rev. B {\bf{60}}, 13470 (1999).

\bibitem{Fisch03:67} R. Fisch, Phys. Rev. B {\bf{67}}, 094110 (2003).

\bibitem{Dkhil04:65} B. Dkhil, J.M. Kiat, G. Calvarin, G. Baldinozzi, S.B. Vakhrushev, S. Suard {\bf{65}}, 023104 (2004).

\bibitem{Xu04:70_1} G. Xu, Z. Zhong, Y. Bing, Z.-G. Ye, C. Stock, and G. Shirane Phys. Rev. B 70, 064107 (2004).

\bibitem{Luo00:39} H. Luo, G. Xu, H. Xu, P. Wang and Z. Yin, Jpn. J. Appl. Phys. {\bf{39}}, 5581 (2000).

\bibitem{Shirane:book} G. Shirane, S.M. Shapiro, and J.M. Tranquada, \textit{Neutron Scattering with a Triple Axis Spectrometer} (Cambridge University Press, Cambridge, 2002).

\bibitem{Gehring04:16} P.M. Gehring, W. Chen, Z.-G. Ye, and G. Shirane, J. Phys.: Condens Matter {\bf{16}}, 7113 (2004).

\bibitem{Shirane51:6} G. Shirane and S. Hoshino, J. Phys. Soc. Jpn. {\bf{6}}, 265 (1951). 

\bibitem{Xu04:70} G. Xu, Z. Zhong, H. Hiraka, and G. Shirane, Phys. Rev. B {\bf{70}}, 174109 (2004).

\bibitem{Hiraka04:70} H. Hiraka, S.-H. Lee, P.M. Gehring, G. Xu, and G. Shirane, Phys. Rev. B {\bf{70}}, 184105 (2004).

\bibitem{Burton99:60} B.P. Burton and E. Cockayne, Phys. Rev. B {\bf{60}}, R12542 (1999).

\bibitem{Laguta03:67} V.V. Laguta, M.D. Glinchuk, S.N. Nokhrin, I.P. Bykov, R. Blinc, A. Gregorovic, and B. Zalar, Phys. Rev. B {\bf{67}}, 104106 (2003).

\bibitem{Boulesteix94:108} C. Boulesteix, F. Varnier, A. Llebaria, and E. Husson, J. Solid State Chem. {\bf{108}}, 141 (1994).

\bibitem{Svitelskiy03:68} O. Svitelskiy, J. Toulouse, G. Young, Z.-G. Ye, Phys. Rev. B {\bf{68}}, 104107 (2003).

\bibitem{Halperin76:14} B.I. Halperin and C.M. Varma, Phys. Rev. B {\bf{14}}, 4030 (1976).

\bibitem{Hirota02:65} K. Hirota, Z.-G. Ye, S. Wakimoto, P.M. Gehring, and G. Shirane, Phys. Rev. B {\bf{66}}, 104105 (2002).

\bibitem{Vak89:90} S.B. Vakhrushev, A. A. Naberezhnov, N.M. Okuneva, and B.N. Savenko, Phys. Solid State {\bf{37}}, 1993 (1995).

\bibitem{Stock04:xx} C. Stock, H. Luo, D. Viehland, J.F. Li, I. Swainson, R.J. Birgeneau, G. Shirane, J. Phys. Soc. Jpn. {\bf{74}}, 3002(2005).

\bibitem{Powell75:12} B.M. Powell and M. Nielsen, Phys. Rev. B {\bf{12}}, 5959 (1975).

\bibitem{Kambaxx:04} S. Kamba, M. Kempa, V. Bovtun, J. Petzelt, K. Brinkman, and N. Setter, cond-mat/0412017.

\bibitem{Kambaxx:05} S. Kamba, M. Berta, M. Kempa, J. Petzelt, K. Brinkman, and N. Setter, cond-mat/0504755.

\bibitem{Yamadaxx:9573} Y. Yamada and T. Takakura, cond-mat/0209573.

\bibitem{Feng96:55} Q. Feng, Q.J. Harris, R.J. Birgeneau, and J.P. Hill, Phys. Rev. B {\bf{55}}, 370 (1997).
\end{document}